\documentstyle[12pt]{article}

\def\be{\begin{equation}}
\def\ee{\end{equation}}
\begin{document}

\title{Strong coupling in brane-induced gravity in
five dimensions}            

\author{
      V.~A.~Rubakov\\
    {\small \em Institute for Nuclear Research of the Russian Academy of
  Sciences,}\\
  {\small \em 60th October Anniversary prospect 7a, Moscow 117312}\\
  }

\maketitle

\begin{abstract}
Brane-induced gravity in five dimensions (Dvali--Gabadadze--Porrati
model) exhibits modification of gravity at ultra-large distances,
$r\gg r_c = M_{Pl}^2/M^3$  where
$M$ is the five-dimensional gravity scale.
This makes the model 
potentially interesting for explaining the observed
acceleration of the Universe. We argue, however,
that it has an intrinsic intermediate
energy scale $(M^9/M_{Pl}^4)^{1/5}$.
At higher energies,
the model is strongly coupled. For $r_c$ of order of the present
Hubble size, the strong coupling regime occurs at distanced below
tens of metres.
\end{abstract}

\section{Introduction}

Whether there exists a consistent and phenomenologically
acceptable theory of gravity in which gravitational interactions get
modified at ultra-large distances is an interesting problem.
Indeed, the large-scale modification of gravity might become an
appealing way of explaining the accelerated expansion of the Universe
at the present epoch. At first sight, modification of gravity at
ultra-large distances may naturally occur in theories with large
or infinite extra dimensions. However, several models of this
sort~\cite{Charmousis:1999rg,Kogan:1999wc,Dvali:2001ae}, in which
linearized gravity experienced by brane matter has purely tensor
structure, have been shown to have
ghosts~\cite{Pilo:2000et,Dubovsky:2002jm}.
Because of the van~Dam--Veltman--Zakharov phenomenon, 
an alternative is that gravity linearized about flat background has a
scalar component. This feature is similar to 4d gravity with massive
graviton, and it is indeed inherent in the five-dimensional induced
gravity model of Dvali, Gabadadze and Porrati
(DGP)~\cite{DGP}, as well as in other no-ghost brane-world
models~\cite{Kogan:2001yr}.

Four-dimensional gravity with massive graviton has been shown to
possess a strong-interaction energy scale, which is intermediate
between the Planck mass and graviton
mass~\cite{Arkani-Hamed:2002sp}. This property, and 
other arguments~\cite{Porrati:2002cp} suggest that similar strong
interaction scale may be present in no-ghost brane-world models with
gravity modified at ultra-large scales. In this note we address this
issue in the framework of DGP model, whose action is a sum
\be
 S_{tot} = S_{bulk} + S_{brane}
\ee
where the 5d bulk piece is
\be
 S_{bulk} = M^3 \int~d^5X~\sqrt{g^{(5)}} R^{(5)}
+ \; \mbox{total~divergence}
\ee
and the brane term is
\be
   S_{brane} = M_{Pl}^2\int_{brane}~d^4x~\sqrt{g^{(4)}} R^{(4)}
\ee
At distances
\be
         M^{-1} \ll r \ll r_c
\ee
where
\be
   r_c = \frac{M_{Pl}^2}{M^3}
\label{i+}
\ee
linearized gravity with sources on the brane is four-dimensional, while
it becomes five-dimensional at ultra-large distances, $r \gg r_c$. 

In this paper we argue that DGP model has an inherent energy scale
\be
 E_{strong} = \left(\frac{M^9}{M^4_{Pl}}\right)^{\frac{1}{5}}
\label{i*}
\ee
At energies above this scale, scalar (in 4d sense) degrees of freedom
become strongly interacting, so the model does not admit classical
treatment. It is worth noting that these degrees of freedom do not
decouple. This makes DGP model an unlikely candidate for explaining
the acceleration of the Universe: if $r_c$ is of the order of the
present Hubble size, the theory is strongly coupled at distances below
$E_{strong}^{-1} \sim (r_c^3 l_{Pl}^2)^{1/5} \sim 30$~m.

This issue has been addressed independently by Luty, Porrati and
Rattazzi~\cite{new}. Their conclusion is similar in spirit, but not in 
detail: their energy scale of strong interaction is $M^2/M_{Pl}$,
which is lower than our scale (\ref{i*}). The origin of this
disagreement is unclear at the moment.

This paper is organized as follows. To get an idea of ``large''
degrees of freedom, we study DGP model in Section 2 at linearized
level (quadratic action). We first find that the full propagator in
de~Donder--Fock gauge has large terms with scalar structure (in 4d
sense). These terms are pure gauge in the bulk but not on the brane. 
We then discuss the quadratic action in Gaussian normal gauge, and
introduce a change of variables after which the degrees of freedom
decouple into tensor part, suppressed near the brane, and unsuppressed
scalar (and vector) part. This change of variables induces large 
brane bending term into the metric, which is again pure gauge in the
bulk but not on the brane. This term is enhanced by $M_{Pl}^2$,
signalizing strong interaction between the scalar modes. In Section 3
we proceed to study the action at cubic order to see that naive power
counting does not work, due to cancellations, but there remain
unavoidable cubic terms enhanced by $M_{Pl}^2$. These are precisely
the terms that make the theory strongly coupled at the energy scale 
(\ref{i*}). We conclude in Section 4.

\section{Linearized theory}

Let us begin with the linearized theory about flat background in 
de~Donder--Fock gauge. It is straightforward to calculate the full
propagator in this gauge. Let
  $D_5 (p;y,y')$
denote the free 5-dimensional  propagator in mixed 
momentum-coordinate representation,
\be
  D_5 = \frac{e^{-p |y-y'|}}{p}
\ee
Hereafter $X^A = (x^\mu, y)$, $A=0,1,2,3,5$ and $p$ is the
four-momentum
(we work with Euclidean version of DGP model).
Let us denote
\be
  D_0 = D_0 (p,y) \equiv D_5 (p;y,0) = \frac{e^{-p |y|}}{p}
\ee
and
\be
  D_0^{\prime} =D_0^{\prime} (p,y') \equiv D_5 (p;0,y')= D_0(p,y')
\ee
We also use the notation 
\be
  D_{0 0} = D_{00} (p) \equiv D_5 (p;0,0) = \frac{1}{p}
\ee
Then the
graviton propagator in DGP model in 
de~Donder--Fock gauge has the following non-zero
components 
\be
D_{5~5}^{5~5} (p, y,y') = \frac{1}{M^3}
\frac{2}{3} D_5 + 
\frac{M_{Pl}^2}{M^6} \frac{2}{3} p^2 D_0 D_0^{\prime}
\label{55}
\ee
\be
 D_{\mu 5}^{5 \nu} = \frac{1}{M^3} D_5 \eta^\nu_\mu
\label{5mu}
\ee
and
\begin{eqnarray}
G^{\nu \rho}_{\mu \lambda}
&=& \frac{1}{M^3} D_5 \left[\frac{1}{2} 
\left( \eta^{\nu \rho} \eta_{\mu \lambda}
+ \eta^{\nu}_{\lambda} \eta_{\mu}^{\rho}\right)
- \frac{1}{3} \eta^{\nu}_{\mu} \eta^{\rho}_{\lambda} \right]
\nonumber \\
&+& \frac{M_{Pl}^2}{M^3} \frac{1}{M^3 + M_{Pl}^2 p^2 D_{00}}
\cdot D_0 D_0^{\prime}
\nonumber \\
&\times& \left[
-p^2 \left( \frac{1}{2} \eta^{\nu \rho} \eta_{\mu \lambda}
+ \frac{1}{2}\eta^{\nu}_{\lambda} \eta_{\mu}^{\rho}
- \frac{1}{3} \eta^{\nu}_{\mu} \eta^{\rho}_{\lambda} \right) \right.
\nonumber \\
&+& \left. \frac{1}{2}(p_\mu p^\rho \eta^\nu_\lambda
+ p^\nu p^\rho \eta_{\mu \lambda} + (\lambda \leftrightarrow \rho))
\right.
\nonumber \\
&-& \left. \frac{1}{3} p_\mu p^\nu \eta^\rho_\lambda
- \frac{1}{3} \eta^\nu_\mu p^\rho p_\lambda
-\frac{2}{3} \frac{p_\mu p^\nu p^\rho p_\lambda}{p^2} \right]
\nonumber \\
&+& \frac{M_{Pl}^2}{M^6} \frac{2}{3}
\frac{p_\mu p^\nu p^\rho p_\lambda}{p^2} \cdot D_0 D_0^{\prime}
\label{munu}
\end{eqnarray}
To make contact with Ref.~\cite{DGP}, one notices that the
brane-to-brane propagator may be written as follows,
\begin{eqnarray}
G^{\nu \rho}_{\mu \lambda}(y=y'=0) &=&
\frac{D_{00}}{M^3 + M_{Pl}^2p^2 D_{00}} 
\left[\frac{1}{2} 
\left( \eta^{\nu \rho} \eta_{\mu \lambda}
+ \eta^{\nu}_{\lambda} \eta_{\mu}^{\rho}\right)
- \frac{1}{3} \eta^{\nu}_{\mu} \eta^{\rho}_{\lambda} \right]
\nonumber \\
&+& \frac{M_{Pl}^2 D_{00}^2}{M^3 (M^3 + M_{Pl}^2p^2D_{00})}
\left[ \frac{1}{2}(p_\mu p^\rho \eta^\nu_\lambda
+ p^\nu p^\rho \eta_{\mu \lambda} + (\lambda \leftrightarrow \rho))
\right.
\nonumber \\
&-& \left. \frac{1}{3} p_\mu p^\nu \eta^\rho_\lambda
- \frac{1}{3} \eta^\nu_\mu p^\rho p_\lambda
-\frac{2}{3} \frac{p_\mu p^\nu p^\rho p_\lambda}{p^2} \right]
\nonumber \\
&+& \frac{M_{Pl}^2}{M^6} \frac{2}{3}
\frac{p_\mu p^\nu p^\rho p_\lambda}{p^2} \cdot D_{00}^2
\end{eqnarray}
The first term here determines the interaction between 
conserved sources on the brane, at linearized level. At intermediate
distances (\ref{i+}) one has $M_{Pl}^2 p^2 D_{00} \gg M^3$,
so this interaction has 4d form. 
Note that the second term 
does not vanish when contracted with conserved $T_\rho^\lambda$, so
matter on the brane couples to scalar degrees of freedom at strength
set by the 5d mass $M$.

Let us come back to the full propagator. It
has large parts, the last terms in
(\ref{55}) and (\ref{munu}). These terms may be gauged away
everywhere in the bulk, but not on the brane. Indeed, they may be
parametrized by introducing a 4d scalar ``field''
$\varphi (x)$ whose propagator equals $1/p^2$, and whose contribution 
to metric is
\begin{eqnarray}
  h_{\mu \nu} (p,y) &=& \sqrt{\frac{2}{3}} \frac{M_{Pl}}{M^3}
p_\mu p_\nu D_0 (p,y)\cdot \varphi(p)
\nonumber  \\
  h_{55} (p,y) &=& \sqrt{\frac{2}{3}} \frac{M_{Pl}}{M^3}
p^2 D_0 (p,y)\cdot \varphi(p)
\label{hphi}
\end{eqnarray} 
Outside the brane, the ``field'' $\varphi$ may indeed be
gauged away. Thus, we see that 4d scalars in DGP model have fairly
peculiar properties; in particular, they appear enhanced by $M_{Pl}$.

To study the model in more detail, let us move to the Gaussian 
normal gauge, 
\be
   h_{55} = h_{5\mu} = 0
\label{GN}
\ee
and calculate the quadratic action.
Let us decompose the metric
\be
  h_{\mu \nu} = h_{\mu \nu}^{TT} + (p_{\mu} u_\nu + p_\nu u_\mu)
+ p_\mu p_\nu v + \frac{1}{2}\eta_{\mu \nu} \phi 
\label{decompose}
\ee
where $h_{\mu \nu}^{TT} $ is transverse traceless (in 4d sense),
\be
   p_\mu h^{TT~\mu}_\nu = h^{TT~\mu}_\mu = 0
\ee
and $u_\mu$ is transverse
\be
   p_\mu u^\mu = 0
\ee
Then one finds, at quadratic order,
\be
  S = \int~dy~d^4 x~ L_{GN}
\ee
where
\begin{eqnarray}
  L_{GN} &=& h^{TT~\mu \nu}[M^3  (\partial_y^2 - p^2 ) - 
M_{Pl}^2 p^2 \delta (y) ] h_{\mu \nu}^{TT}
\nonumber \\
&+& 2M^3 u^\mu p^2 \partial_y^2 u_\mu
\nonumber \\
 &+& 3[M^3  \phi p^2 \partial_y^2 v +  
M^3 \phi \partial_y^2 \phi
 - \frac{M^3}{2} \phi p^2 \phi - \frac{M_{Pl}^2}{2} \delta (y) 
\phi p^2 \phi]
\label{L}
\end{eqnarray}
We are interested in the part that
contains 4d scalars $v$ and $w$, last line in eq.~(\ref{L}). 
We can get rid of the very last term, which is proportional to
$M_{Pl}^2$, by defining a new field $\hat{v}$, such 
that
\be
   v (p,y) = \phi(p,0) \cdot \frac{M_{Pl}^2}{2M^3} |y| + 
\hat{v}(p,y)
\label{iii*}
\ee
Then in tems of $w$ and $\hat{v}$ the last line in 
(\ref{L})  is precisely the part of quadratic 5d action
(in the gauge (\ref{GN})) that contains 4d scalars,
\be
 L_{scalar} = 3M^3  (\phi p^2 \partial_y^2 \hat{v}  +  \phi 
\partial_y^2 \phi
 - \frac{1}{2}\phi p^2 \phi)
\ee
Hence, ``canonically normalized'' scalars and vectors are
\be
   \hat{v}^{can}, \phi^{can}, u_{\mu}^{can} = 
\frac{\hat{v}}{M^{3/2}} \; , \;\;\;\;
\frac{\phi}{M^{3/2}}  \; , \;\;\;\; \frac{u_\mu}{M^{3/2}}
\label{canon}
\ee
where we made use of the fact that
the quadratic term with $u_\mu$, second line in
(\ref{L}), also has the 5d form.

But the metric contains large piece
\be
   h_{\mu \nu}^{large} = \partial_{\mu} \partial_{\nu}
\frac{M_{Pl}^2}{2M^3} |y|\cdot \phi (p,0)
\label{large}
\ee
This piece is pure gauge everywhere outside the brane.
Say, at $y>0$ the large piece may be gauged away by the gauge
transformation with
\begin{eqnarray}
   \xi_5 &=& \frac{M_{Pl}^2}{M^3}  \cdot \phi (p,0)
\nonumber \\
   \xi_\mu &=& - \frac{M_{Pl}^2}{M^3} \cdot y \cdot \phi (p,0)
\end{eqnarray}
On the other hand, 
the large piece (\ref{large}) is not pure gauge on the brane.

Most naively, the cubic and higher order terms in the action
appear to be enhanced by high powers of $M_{Pl}^2$, because of the
presence of the large piece (\ref{large}) in the metric. However,
since this piece is longitudinal in 4d sense, and pure gauge outside
the brane, one expects strong cancellations. Let us see that 
large terms in the 
cubic action indeed cancel, but not completely.

\section{Cubic order}

We still work in Gaussian normal gauge,
\be
   g_{55} = 1 \;,\;\; g_{5 \mu} = 0
\label{gauge-cub}
\ee
The brane is placed at $y=0$, and we consider metric $g_{\mu \nu}$
symmetric with respect to the brane,
\be
  g_{\mu \nu}(x, -y) = g_{\mu \nu}(x,y)
\label{symmetry-cub}
\ee 
The bulk Lagrangian may be conveniently written in the
 form
\begin{eqnarray}
L_{bulk} &=& M^3 \sqrt{g} R^{(4)} +
M^3 \sqrt{g} \left(\frac{1}{4} g^{\mu \nu} \partial_5 g_{\mu \nu}
          g^{\lambda \rho} \partial_5 g_{\lambda \rho}
- \frac{1}{4} g^{\mu \nu} g^{\lambda \rho}
\partial_5 g_{\mu \lambda } \partial_5 g_{\nu \rho} \right)
\nonumber \\
&\equiv& M^3 \sqrt{g} R^{(4)} + \Delta L
\label{ii*}
\end{eqnarray}
We pursue the idea of making the change of variables from the metric 
$g_{\mu \nu}$ to another ``metric'' $\hat{g}_{\mu\nu}$ in such 
a way that the largest pieces of the brane action are cancelled by
the contribution due to the bulk action.
The metric $\hat{g}_{\mu \nu}$ is related to $g_{\mu \nu}$ by a gauge
transformation on the right of the brane, and another gauge
transformation on the left of the brane, so new contribution
to the action
appears at the brane only. The transformation from
metric $g_{\mu \nu}$ to $\hat{g}_{\mu \nu}$ is single-valued on the
brane, so  metric induced on the brane is uniquely defined in terms
of $\hat{g}_{\mu \nu}$. Metric $\hat{g}_{AB}$ still has to obey
the gauge conditions (\ref{gauge-cub}) and symmetry property
(\ref{symmetry-cub}).

To this end, let us study what are the 5d gauge 
transformations that leave the conditions (\ref{gauge-cub})
satisfied. We write
\be
  g_{AB} (X) = \frac{\partial \hat{X}^C}{\partial X^A} 
               \frac{\partial \hat{X}^D}{\partial X^B}
               \hat{g}_{CD}(\hat{X}(X))
\ee
where $X^A = (x^\mu,y)$, and
in perturbation theory 
\be
   \hat{X}^A = X^A + \xi^A (X)
\ee
We will need the relation between metric perturbations 
$h_{\mu \nu} $ and $\hat{h}_{\mu\nu}$
at quadratic order,
\be
 h_{AB} = \partial_A \xi_B + \partial_B \xi_A
+ \partial_C \hat{h}_{AB} \xi^C
+\partial_A \xi^C \hat{h}_{BC} 
+\partial_B \xi^C \hat{h}_{AC} 
+ \partial_A \xi^C \partial_B \xi_C
\ee
where all functions are functions of $X$ and indices are raised and
lowered by  Euclidean metric.
 We require
\be
  \hat{h}_{55} = 0
\ee
on the left of the brane, and on the right of the brane separately.
This gives the following equation
\be
  2 \partial_5 \xi_5 + \partial_5 \xi^5 \cdot \partial_5 \xi_5
  + \partial_5 \xi^\mu \cdot \partial_5 \xi_\mu = 0
\label{55-cub}
\ee
The requirement
\be
 \hat{h}_{5\mu} = 0
\ee
gives
\be
\partial_5 \xi_\mu + \partial_\mu \xi_5
+ \partial_5 \xi^5 \cdot \partial_\mu \xi_5
+ \partial_5 \xi^\nu \cdot \partial_\mu \xi_\nu
+ \partial_5 \xi^\nu \cdot \hat{h}_{\mu \nu} = 0
\label{5mu-cub}
\ee
We stress that these equations should be satisfied on the left of the
brane and on the right of the brane separately. 
To ensure the symmetry property (\ref{symmetry-cub})
for both  $h_{\mu \nu}$ and  $\hat{h}_{\mu \nu}$, we take
$\hat{X}^\mu$  symmetric, and
$\hat{X}^5$  anti-symmetric in $y\equiv X^5$, that is
\begin{eqnarray}
    \xi^\mu (x,-y) &=& \xi^\mu (x,y)
\nonumber \\
    \xi^5 (x,-y) &=& - \xi^5 (x,y)
\label{sym-bis-cub}
\end{eqnarray}
Solving eqs. (\ref{55-cub}) and (\ref{5mu-cub}) on the left 
and on the right of the brane, and imposing (\ref{sym-bis-cub}),
we find
\be
 \xi_5 = \epsilon \cdot \mbox{sign} (y) - 
\frac{1}{2}\partial_\mu \epsilon \partial^\mu \epsilon
 \cdot y  
\label{xi5-cub}
\ee
\be
\xi_\mu = - \partial_\mu \epsilon \cdot |y|
          + \partial_\nu \epsilon \cdot \mbox{sign} (y) \cdot 
\int_0^y~dy' \hat{h}_\mu^\nu
\label{ximu-cub}
\ee
where $\epsilon$ is an arbitrary 
function of 4d coordinates only,
\be
    \epsilon = \epsilon (x^\mu)
\ee
Physically, $\epsilon (x)$ is a 
brane bending function, at least to linear
order in metric perturbations.
We will use the freedom parametrized by $\epsilon (x)$
to get rid of certain large terms in the total action.

The 4d components of the metric are then
\begin{eqnarray}
   h_{\mu \nu} = \hat{h}_{\mu \nu}
   &+& \partial_\mu \xi_\nu + \partial_\nu \xi_\mu
 + \partial_\mu \xi^5 \partial_\nu \xi_5 
 + \partial_\mu \xi^\lambda \partial_\nu \xi_\lambda
\nonumber \\
 &+&  \partial_\mu \xi^\lambda \hat{h}_{\nu \lambda}
 + \partial_\nu \xi^\lambda \hat{h}_{\mu \lambda}
 + \partial_5 \hat{h}_{\mu \nu} \xi^5 + \partial_\lambda 
\hat{h}_{\mu \nu} \xi^\lambda
\label{tmp55-cub}
\end{eqnarray}
Note that in spite of the jump of $\xi^5$, the latter relation
is well defined, i.e., $h_{\mu \nu}$ is uniquely defined on the brane
in terms of $\hat{h}_{\mu \nu}$ and $\epsilon$ (provided that
$\hat{h}_{\mu \nu}$ is symmetric). 

Let us clarify the logic again. We may forget about previous steps,
and merely consider eq.~(\ref{tmp55-cub}) as the definition
of the change of variables from $h_{\mu \nu}$ to $\hat{h}_{\mu \nu}$ in
the gauge (\ref{gauge-cub}), with $\xi^5$ and $\xi^\mu$ defined by 
eqs.~(\ref{xi5-cub}) and (\ref{ximu-cub}) in terms of yet arbitrary
function $\epsilon (x)$. The further procedure is to calculate the
total action, up to cubic order, and then choose $\epsilon (x)$ to
simplify this action.

We will need the expression
for $h_{\mu \nu}$ on the brane:
\be
  h_{\mu \nu} (y=0) = \hat{h}_{\mu \nu}
 + \partial_\mu \epsilon \partial_\nu \epsilon
\label{ind-cub}
\ee

Let us now calculate the action in terms of $\hat{h}_{\mu \nu}$ and 
$\epsilon$. We begin with the bulk term. Since $h_{\mu \nu}$
does not jump across the brane, the action is the sum of integrals
of the Lagrangian (\ref{ii*}) over regions
left and right of the brane. We write in each of these regions
\be
   (\sqrt{g} R^{(5)}[g])(X) = \mbox{det} 
\left( \frac{\partial \hat{X}^A}{\partial X^B} \right) 
(\sqrt{\hat{g}} R^{(5)}[\hat{g}])(\hat{X}(X))
\ee
This gives, up to cubic order,
\begin{eqnarray}
 \sqrt{g} R^{(5)}[g]
 &=& \sqrt{\hat{g}} \hat{R}^{(5)} + \partial_A (\sqrt{\hat{g}} 
\hat{R}^{(5)}
\xi^A) 
\nonumber \\
&+& \frac{1}{2}
 \partial_A [\partial_B (\sqrt{\hat{g}}\hat{R}^{(5)}) \xi^A \xi^B
+ \sqrt{\hat{g}} \hat{R}^{(5)} \xi^A \partial_B \xi^B
\nonumber \\   
&-& \sqrt{\hat{g}} \hat{R}^{(5)} \xi^B \partial_B \xi^A] 
\label{R-cub}
\end{eqnarray}
where
\be
\hat{R}^{(5)} = R^{(5)}[\hat{g}]
\ee
and all quantities are functions of $X$. 
Since $\hat{g}$ and $\hat{R}^{(5)}$ are symmetric, and $\xi_\mu$
vanishes at the brane, one finds that the
integration of (\ref{R-cub})
over regions left and right of the brane gives the following
additional contribution to the action
\be
  -2 M^3 \int_{brane}~d^4x~\sqrt{\hat{g}} \hat{R}^{(5)} \epsilon
\label{tmp1-cub}
\ee
Let us now consider the second term in eq.(\ref{ii*}).
It is equal to
\be
   2 M^3 \partial_5^2 \sqrt{g}
\ee
and hence contributes to the action as 
\be
 - 2M^3 \left[\partial_5 \sqrt{g}\right]_{y \to -0}^{y \to +0}
\label{gjump-cub}
\ee
Now, we have
\be
  \sqrt{g} = \mbox{det}(\delta^\mu_\nu + \partial_\nu \xi^\mu)
\cdot \sqrt{\hat{g} (x^\mu + \xi^\mu, y + \xi^5)}
\ee
For $\xi^5 =0$
the right hand side is a total 4d divergence, so we 
have to evaluate the determinant here to the first order only,
$ \mbox{det}(\delta^\mu_\nu + \partial_\nu \xi^\mu)
= 1 + \partial_\mu \xi^\mu$.  
Then modulo total 4d divergence, one has up to cubic order
\be
\sqrt{g} = \sqrt{\hat{g}} +
\mbox{det}(\delta^\mu_\nu + \partial_\nu \xi^\mu)
\cdot
\left(\partial_5 \sqrt{\hat{g}} \cdot \xi^5
+ \partial_\sigma \partial_5 \sqrt{\hat{g}} \cdot
\xi^\sigma \xi^5
+ \frac{1}{2} \partial^2_5 \sqrt{\hat{g}} \cdot
\xi^5 \xi^5
\right)
\ee
which gives, again up to total 4d divergence,
\be
\sqrt{g} = \sqrt{\hat{g}}
+ \partial_5 \sqrt{\hat{g}} \cdot \xi^5
- \partial_5 \sqrt{\hat{g}} \cdot \xi^\mu \partial_\mu \xi^5
+  \frac{1}{2} \partial^2_5 \sqrt{\hat{g}} \cdot \xi^5 \xi^5
\ee
Now, both $\partial_5 \sqrt{\hat{g}}$ and $\xi^\mu$ vanish on the
brane, so the third term here does not contribute
to (\ref{gjump-cub}). The fourth term does not contribute at
cubic level too, because $(\xi^5)^2 = \epsilon^2$ is continuous
at quadratic level. The contribution from the second term is entirely
due to the first term in (\ref{xi5-cub}), since the second term 
in (\ref{xi5-cub}) is 
smooth across the brane. Thus, the additional
contribution to the action is
\be
 - 4 M^3 \int_{brane}~d^4x ~\partial^2_5\sqrt{\hat{g}} \cdot \epsilon
\ee
This contribution, together with the contribution (\ref{tmp1-cub})
adds to
\be
  - 2 \int_{brane}~d^4x~ \epsilon \cdot L_{bulk}[\hat{g}]
\equiv -2  \int_{brane}~d^4x~ \epsilon (x) \cdot 
(M^3 \sqrt{\hat{g}}\hat{R}^{(4)}  
+ \Delta L [\hat{g}_{\mu\nu}])
\label{bulk-brane-cub}
\ee
where $\Delta L$ is defined in eq.~(\ref{ii*}).

Let us  turn to the brane action.
The contribution due to $\epsilon$ into the brane action comes from
the second term in eq.~(\ref{ind-cub}). To cubic order, it is
\be
        M_{Pl}^2 \int~d^4x \sqrt{\hat{g}}
\left (\hat{R}^{(4)~\mu \nu} - \frac{1}{2} \hat{g}^{\mu \nu} 
\hat{R}^{(4)}
\right) \partial_\mu \epsilon \partial_\nu \epsilon
\label{brane-brane-cub}
\ee
Thus, the total action $S_{tot}[g]$ equals $S_{tot}[\hat{g}]$ 
plus the sum of
(\ref{bulk-brane-cub}) and (\ref{brane-brane-cub}).

Now, let us discuss scalar and vector modes (in 4d sense).
These are parametrized as follows,
\be
\hat{g}_{\mu \nu} (x,y)= 
\frac{\partial \tilde{x}^\lambda}{\partial x^{\mu}}
\frac{\partial \tilde{x}^\rho}{\partial x^{\nu}}
\tilde{g}_{\lambda \rho} (\tilde{x}(x,y), y)
\ee
where
\be
   \tilde{x}^\mu = x^\mu + \pi^\mu (x,y)
\ee
(at linear level, $\pi_\mu = u_\mu + \partial_\mu v$
in notations of section 2) 
and
\be
 \tilde{g}_{\mu \nu} = \mbox{e}^{2\phi (x,y)} \eta_{\mu \nu}
\ee
The brane action is then
\be
   S_{brane}[\hat{g}] = 6 M_{Pl}^2 \int~d^4x~ \mbox{e}^{2\phi}
   \partial_\mu \phi \partial^\mu \phi
\ee
Thus, the total action, up to cubic level, is
\be
  S_{tot} = S_{bulk} [\hat{g}]  + S_{brane}[\hat{g}]
      + S_{\epsilon}
\ee
where $S_{\epsilon}$ is the additional brane term, 
the sum of
(\ref{bulk-brane-cub}) and (\ref{brane-brane-cub}).
We write
\be
 \hat{R}^{(4)}_{\mu \nu} - \frac{1}{2} \hat{g}_{\mu \nu} \hat{R}^{(4)} = 
-2 \partial_\mu \partial_\nu \phi + 
2 \eta_{\mu \nu} \Box^{(4)} \phi + \; (\mbox{higher~orders~in~}
\phi, \pi)
\ee
and
\be
    \sqrt{\hat{g}} \hat{R}^{(4)}
=
-6 \Box^{(4)} \phi   + \; (\mbox{higher~orders~in~}
\phi, \pi)
\ee
and obtain explicitly
\begin{eqnarray}
   S_{\epsilon} = \int~d^4x~\left[
    12M^3 \epsilon \cdot  \Box^{(4)} \phi 
   \right. &+& 
\left. M_{Pl}^2 \cdot \partial_\mu \epsilon \partial_\nu \epsilon
 \cdot (-2 \partial_\mu \phi + 2 \eta_{\mu \nu} \Box^{(4)} \phi)
\right.
\nonumber \\   
  &+& \left. 2M^3 \epsilon \cdot {\cal L}_2 (\phi, \pi^\mu) \right]
\end{eqnarray}
where ${\cal L}_2$ is quadratic in fields $\phi$ and $\pi_\mu$,
\begin{eqnarray}
{\cal L}_2 = &-&12 \phi \Box^{(4)}\phi -
6 \partial_\mu \phi \partial^\mu \phi
-6 \partial_\mu (\Box^{(4)} \phi \cdot \pi^\mu)
+ 12 (\partial_5 \phi)^2 + 6 \partial_5 \phi \partial_5\partial_\mu \pi^\mu 
\nonumber \\
&+& (\partial_5 \partial_\mu \pi^\mu)^2
-\frac{1}{2} \partial_\mu \partial_5 \pi_\nu
 \partial^\nu \partial_5 \pi^\mu
-\frac{1}{2} \partial_\mu \partial_5 \pi_\nu
 \partial^\mu \partial_5 \pi^\nu
\label{ii+}
\end{eqnarray}
At quadratic order, the relevant terms in the action are
\be
  6 \int~d^4x~ (2M^3 \epsilon^{(1)} \Box^{(4)} \phi + M_{Pl}^2
  \partial_\mu \phi \partial^\mu \phi)
\label{tmp12}
\ee
where $\epsilon^{(1)}$ is linear in metric perturbations.
To get rid of these terms, we choose
(cf. eq.~(\ref{iii*}))
\be
  \epsilon^{(1)} (x) = \frac{M_{Pl}^2}{2M^3} \cdot \phi (x, y=0)
\ee
then the contribution (\ref{tmp12}) vanishes.

At cubic level, the largest terms are
\begin{eqnarray}
 \int~d^4x &&  [12 M^3 \epsilon^{(2)} \cdot \Box^{(4)} \phi
+ M_{Pl}^2 \partial^\mu \epsilon^{(1)} \cdot \partial^\nu
\epsilon^{(1)} \cdot
(-2 \partial_\mu \partial_\nu \phi + 
2 \eta_{\mu \nu} \Box^{(4)} \phi)] 
\nonumber \\
= \int~d^4x && \left[ 12 M^3 \epsilon^{(2)} \cdot \Box^{(4)} \phi
\right.
\nonumber \\
&& + \left. \frac{M_{Pl}^6}{4 M^6} \cdot \partial^\mu \phi  \partial^\nu
\phi \cdot
(-2 \partial_\mu \partial_\nu \phi + 
2 \eta_{\mu \nu} \Box^{(4)} \phi ) \right] 
\end{eqnarray}
These are cancelled out by choosing
\be
\epsilon^{(2)} (x) = \frac{M_{Pl}^6}{12 M^9} \cdot
(\partial_\mu \phi \partial^\mu \phi ) (x, y=0)
\label{E2-cubic}
\ee
After this choice is made, the action for 4d scalars
and vectors becomes, up to cubic level,
\be
S_{tot} = S_{bulk} (\phi, \pi^\mu)
+ M_{Pl}^2 \int_{brane} ~d^4x~\phi \cdot {\cal L}_2 (\phi, \pi^\mu)
\ee
Thus, we see that due to the cancellations, the largest possible terms
disappear, and the cubic action is enhanced by $M_{Pl}^2$ only.

In fact, we can do better by choosing
\be
\epsilon^{(2)} (x) = \frac{M_{Pl}^6}{12 M^9} \cdot
(\partial_\mu \phi \partial^\mu \phi ) (x, y=0)
+ \frac{M_{Pl}^2}{4M^3}(3\phi^2 - 2 \partial_\mu \phi \pi^\mu)
\ee
Then those terms in eq.~(\ref{ii+}) that do not contain transverse
derivatives, cancel out, and we obtain finally
\begin{eqnarray}
S_{tot} &=& S_{bulk} (\phi, \pi^\mu)
+ M_{Pl}^2 \int_{brane} ~d^4x~\phi \cdot
\left[ 12 (\partial_5 \phi)^2 
+ 6 \partial_5 \phi \partial_5\partial_\mu \pi^\mu \right.
\nonumber \\
&+& \left. (\partial_5 \partial_\mu \pi^\mu)^2
-\frac{1}{2} \partial_\mu \partial_5 \pi_\nu
 \partial^\nu \partial_5 \pi^\mu
-\frac{1}{2} \partial_\mu \partial_5 \pi_\nu
 \partial^\mu \partial_5 \pi^\nu \right]
\end{eqnarray}
Further reduction of the cubic action is impossible, as the remaining
terms are not proportional to $\Box^{(4)} \phi$.

With ``canonicaly normalized'' scalars and vectors
(\ref{canon}), the largest interaction term is
proportional to
\be 
 \frac{M_{Pl}^2}{M^{9/2}}
\ee
On dimensional grounds
we conclude that the energy scale of strong interaction between 
the scalar and vector modes is
indeed given by eq.~(\ref{i*}).

\section{Discussion}

The approach taken in this paper is not quite satisfactory.
The problems with this approach are twofold.
First, it is not at all obvious that the conclusion that
the strong interaction scale is (\ref{i*}) will not be
different when higher (quartic, etc.) orders are included.
Indeed, the powers of $M_{Pl}/M$ proliferate, as is seen in
eq.~(\ref{E2-cubic}). Second, the geometrical meaning of
the whole procedure of cancelling
large terms is obscure. On the other hand, our calculation does
 show that DGP model becomes strongly coupled at low energy scale.
It remains to be understood whether this feature is inevitable in
no-ghost models with modification of gravity
at ultra-large scales.

The author is indebted to A.~Barvinsky, D.~Levkov, R.~Rattazzi,
S.~Sibiryakov and P.~Tinyakov for many stimulating discussions.

\end{document}